\documentclass[a4paper]{jpconf}
\usepackage{graphicx}
\usepackage{amsbsy}
\usepackage{color}
\usepackage{amsmath}
\newcommand{\localDM}{\rho_{\mathrm{DM},\odot}}
\begin{document}
\title{Dark matter local density determination based on recent observations}

\author{Pablo F. de Salas}

\address{The Oskar Klein Centre for Cosmoparticle Physics, Department of Physics,
Stockholm University, AlbaNova, 10691 Stockholm, Sweden}

\ead{pablo.fernandez@fysik.su.se}

\begin{abstract}
The local density of dark matter is an important quantity. On the one hand, its value is needed for dark matter direct detection searches. On the other hand, a precise and robust determination of the local dark matter density would help us learn about the shape of the dark matter halo of our Galaxy, which plays an important role in dark matter indirect detection searches, as well as in many studies in astrophysics and cosmology. 
There are different methods available to determine the local dark matter density. Among them, it is common to study either the vertical kinematics of a selected group of tracers or the rotation curve of the Milky Way.
Recent estimates of the local dark matter density have used the precise observations conducted by the ESA/Gaia mission.
However, in spite of the quality of the data released by Gaia's observations, different analyses of the local dark matter density produce dissimilar results.
After a brief review of the most common methods to estimate the local density of dark matter,
here we argue about different explanations for the discrepancies in the results of recent analyses.
We finish by presenting new approaches that have been proposed in the literature and could help us improve our knowledge of this important quantity.
\end{abstract}

\section{Introduction}
Since the first studies by \cite{Kapteyn:1922zz} and \cite{Jeans_1922}, there has been almost a century of estimates of the local dark matter density ($\localDM$). An accurate determination of this quantity is imperative in order to interpret a future dark matter (DM) signal in direct detection searches, because the value of $\localDM$ is degenerate 
with the cross section of the DM-nucleus interaction.
In addition, knowing $\localDM$ is important for indirect detection DM searches and for those studies in cosmology and astrophysics that depend on the distribution of DM in the Milky Way.

There are different ways of making an estimate of the local DM density, and traditionally they are divided into two categories (see e.g. \cite{Read:2014qva}): \emph{local} methods, those that focus on a small volume around the location of the Earth in the Galaxy, and \emph{global} methods, those that instead analyse data that cover a larger volume.
The most common local method is based on the study of the vertical kinematics of stars and it is known as the vertical Jeans equation method, while the most common global method makes use of the rotation curve of the Galaxy.

The precision at which $\localDM$ is known depends on the precision of the observations used to derive its value, as well as on the intrinsic precision of the chosen method. 
Historically, each time new and more accurate data became available, the estimates of $\localDM$ became compatible within a closer range of values, with the increasing precision limited by the assumptions made in the analyses. 
This tendency is easy to appreciate in Fig. 2 of \cite{Read:2014qva} and in Fig. 1 of \cite{Xia:2015agz}. 
Interestingly, the situation has however not converged with the incoming of Gaia's observations. 

Since its first data release (DR1) in September 2016 \cite{Brown:2016tnb}, the ESO/Gaia satellite mission has led precision astronomy to a next level. The quality and quantity of Gaia's observations, especially in their second data release (DR2) 
\cite{Brown:2018dum}, has given the community a precise image of the Galaxy, 
that will improve even further with the final release of the Gaia mission.

Based on this data, recent estimates of $\localDM$ have been performed, but the convergence towards a common value of $\localDM$ has not improved accordingly. In fact, larger deviations are found among recent analyses than in previous estimates of $\localDM$.

\section{Methods to estimate $\boldsymbol{\localDM}$}\label{sec:methods}
In order to estimate the local dark matter energy density, $\localDM$, there are several approaches that can be followed. In general, all of them are based on three key aspects:
\begin{enumerate}
\item Choose one or more tracers.
\item Select a theoretical model for the gravitational potential of the Galaxy.
\item Use Boltzmann and Poisson equations to connect both theory and observation.
\end{enumerate}

There are many astrophysical objects that can be used as tracers, from individual stars belonging to a specific stellar population to a collection of gravitationally bound stars that form a globular cluster. The choice depends on the type of analysis that is intended. 
For instance, stars are very useful if one wants to analyse a local region around our location in the Galaxy, or if one wants to study the rotation curve from objects that are close to the Galactic plane, while globular clusters are a better option if one wants to complement the analysis with some features of the Galactic halo. 

The chosen model for the gravitational potential of the Galaxy also depends on the analysis. If the study is focused on a local volume, typical assumptions such as axisymmetry or the Galaxy being in a steady state are less restrictive than when a larger volume is needed in the analysis, as it happens in those studies based on the rotation curve method.

In order to put together the information that is extracted from observations and the variables that are going to be estimated, we need one important equation. Assuming that our tracers are a selection of stellar populations, we need to solve the collisionless Boltzmann equation
\begin{equation}\label{eq:boltzmann}
\frac{\mathrm{d} f}{\mathrm{d} t} = \frac{\partial f}{\partial t} + \nabla_x f \cdot \mathbf{v} - \nabla_v f \cdot \nabla_x \phi = 0,
\end{equation}
where $f(\mathbf{x},\mathbf{v},t)$ is the distribution function of the tracer at a position $\mathbf{x}$, velocity $\mathbf{v}$ and time $t$, and $\phi$ is the gravitational potential. Given that we know the position and velocity of Galactic objects, but we have little or no knowledge of their accelerations, the assumption that the Milky Way is in a steady state is a necessary requirement in the analyses looking for an estimate of $\localDM$. Therefore, the time dependence of the distribution function is dropped in Eq.~\eqref{eq:boltzmann}.
We also need a connection between the gravitational potential of the Galactic components and the shape of their energy density $\rho$. For that purpose we use the Poisson equation
\begin{equation}\label{eq:Poisson}
\nabla^2_x \phi = 4\pi G \rho,
\end{equation}
where $G$ is the Newton's gravitational constant.

\subsection{Distribution function fitting method}
We can fit $f(\mathbf{x},\mathbf{v})$ directly to the data. However, 
solving Boltzmann equations \eqref{eq:boltzmann} 
for the chosen tracers can be very involved, especially since we need to infer the trajectory of a sufficiently large sample of stars for varying parameters of a Galactic model.
The advantage is that we can exploit the information content in discrete data, but we still have to model the functional form of $f(\mathbf{x},\mathbf{v})$.

Because of its expensive computational needs, this method has been more extensively used in the last decade, especially making use of the action-angle variables 
(see e.g. \cite{BinneyTremaine:book,Binney:2015gaa,Sanders:1511.08213,Cole:2016gzv}).
Here we focus however on another two methods more commonly used.

\subsection{Vertical Jeans equation method}
Instead of using directly the Boltzmann equation to analyse the gravitational potential of the Galaxy, we can study the moments of the distribution function. This is done by means of the derived Jeans equations, which are obtained from Eq.~\eqref{eq:boltzmann} by integrating over all velocities after multiplying by the individual velocity components~\cite{Jeans_1922,BinneyTremaine:book}.
Assuming axisymmetry and a steady-state Galaxy, the vertical $z$-Jeans equation in cylindrical coordinates reads
\begin{equation}\label{eq:z-Jeans}
\frac{1}{R} \frac{\partial \left( R \nu \overline{v_{R}v_z} \right)}{\partial R}
+ \frac{\partial \left( \nu \overline{v^2_z} \right)}{\partial z}
+ \nu \frac{\partial \phi}{\partial z}
=0,
\end{equation}
where $\nu$ is the tracer number density, $\overline{v^2_z}$ the average squared vertical velocity and $\frac{1}{R} \frac{\partial \left( R \nu \overline{v_{R}v_z} \right)}{\partial R}$ is called the `tilt' term. 

Because of the presence of the Galactic disc, stars orbiting the Milky Way close to the Galactic plane at $z=0$ oscillate up and down in the vertical direction as they move. Therefore, we can study the vertical kinematics of nearby stars, those that are placed at our location in the Milky Way, and apply
Eq.~\eqref{eq:z-Jeans} together with Eq.~\eqref{eq:Poisson} in order to estimate $\localDM$.

The clear advantage of this method is that we are directly sensitive to the local value of $\rho_{\rm DM}$, so it is enough that the assumption of axisymmetry is fulfilled at a local level. At the same time, we can be sensitive to any local change in density with respect to the average value of $\rho_{\rm DM}$ at a radial distance $R_\odot$ from the Galactic centre. 

On the other hand, an obvious disadvantage comes from the limited size of the studied volume, in particular if the Milky Way is not entirely in a steady state as it will be discussed later in section~\ref{subsec:disequilibria}.

\subsection{Rotation curve method}
As it has been done for many external galaxies since the pioneering work of Vera Rubin~\cite{Rubin:1980zd}, the rotation curve of the Milky Way can also be used to estimate $\localDM$.

With this method, the theoretical circular velocity $v_{\rm c}$ from the definition
\begin{equation}\label{eq:vc-th}
v_{\rm c}^2(R) = \left. R \frac{\partial \phi}{\partial R} \right|_{z=0}
\end{equation}
is compared at different radii with its observational counterpart (see e.g.~\cite{deSalas:2019pee}).

The advantage of this method over the vertical Jeans equation method is the larger volume that is covered, which allows us to average small deviations from a smooth behaviour in the tracer distribution. Of course, this means that the resulting $\localDM$ is going to be the average value at a distance $R_\odot$ from the Galactic centre, while in principle we could be sitting in a place in the Galaxy with a density different from the smooth prediction of the theoretical models.

\section{Recent estimates of $\boldsymbol{\localDM}$}
\begin{figure}
\centering
\includegraphics[width = 0.6\textwidth]{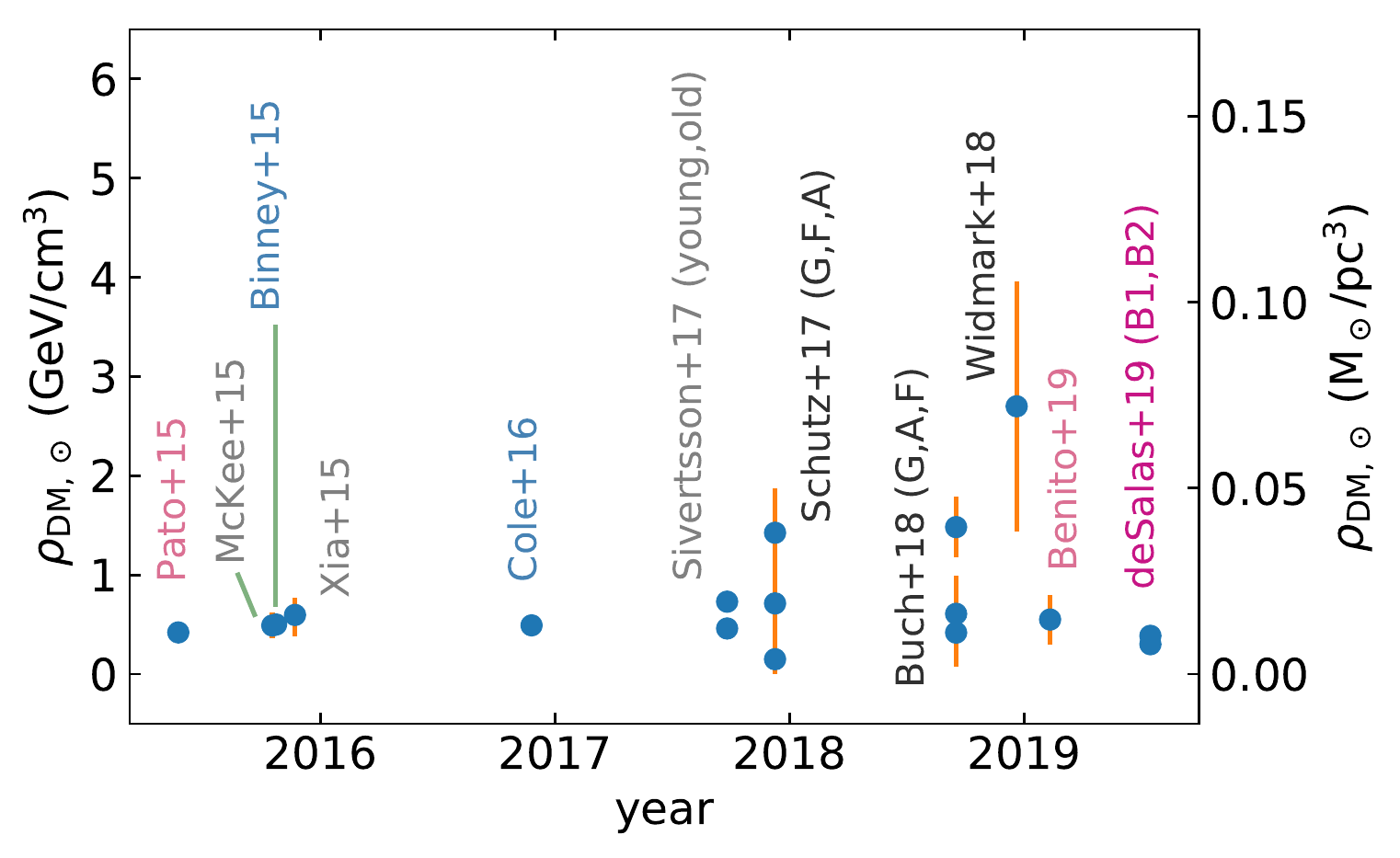}
\caption{Selection of recent studies on $\localDM$ organised according to their publication date. Colours are chosen depending on the method used in the estimate: 
rotation curve (pink), distribution function fitting (blue) and vertical Jeans equation (gray).
In darker colours are the four shown estimates that included Gaia's observations. 
}
\label{fig:rho_vs_years_recent}
\end{figure}

In Fig.~\ref{fig:rho_vs_years_recent} we list a selection of recent $\localDM$ estimates, colour coded according to the method used to obtain the $\localDM$ value. Those studies based on the rotation curve (\cite{Pato:2015dua,Benito:2019ngh,deSalas:2019pee}), the distribution function fitting (\cite{Binney:2015gaa,Cole:2016gzv}) or the vertical Jeans equation method (\cite{McKee:2015hwa,Xia:2015agz,Sivertsson:2017rkp,Schutz:2017tfp,Buch:2018qdr,Widmark:2018ylf}) are presented in pink, blue and gray, respectively.
The colour is darker for the analyses \cite{Schutz:2017tfp,Buch:2018qdr,Widmark:2018ylf,deSalas:2019pee} to show the use of Gaia's observations.

It is easy to see a clear difference in the precision and agreement of earlier $\localDM$ estimates and those performed after 2017, especially in the three estimates that included Gaia's observations and applied the vertical Jeans equation method.
This is surprising, since naively we would expect that the use of better data would lead towards better $\localDM$ estimates.
The increased errorbars in recent works is connected with both a reduced number of assumptions and a smaller volume covered than in previous studies.
On the other hand, there are several causes that could explain why some of the estimates that belong to a common work do not fully agree with each other.
However, a remark should be made about the discrepancies: they look larger than it might have been expected, but most of the estimates are nonetheless statistically compatible with each other at the $2\sigma$ level.
In any case, in the next section we describe possible sources that could explain the differences in the outcomes of recent analyses.

\section{Sources of deviation in the estimated $\boldsymbol{\localDM}$}
There are several reasons why it is expected that different $\localDM$ estimates do not fully agree with each other, and some of these reasons can explain a larger deviation than the inherent differences between the analyses. 

On the one hand, the phase space covered by two distinct tracers, for example two stellar populations with different metallicities, do not need to be the same. The methods described in Sec.~\ref{sec:methods} also lead to estimated $\localDM$ values that do not necessarily agree,
depending mainly on how similar the modelled Galaxy is to the Milky Way.

On the other hand, it is known that the Milky Way is far from equilibrium and there are compelling evidences that show disequilibria in our local neighbourhood. This particular effect, depending on its origin, can cause deviations in different tracer populations even when the same volume and method are used to estimate $\localDM$.

Of course, we can also involve new physics to explain non-coincident $\localDM$ estimates, and we will particularly discuss the implications that the existence of a dark disc 
would have in a comparison between $\localDM$ determinations using the rotation curve method and those based on the vertical kinematics of neighbouring stars.

\subsection{Differences in the data}
Knowing the exact age of a given single star is difficult, but assuming general evolutionary models and provided that we have enough additional information for the star (such as its distance or metallicity), we can compare the colour-magnitude measurement with different theoretical isochrones (assuming different ages and metallicities), from which we can then infer the age of the star.
However, this process becomes simpler if we have a population of stars 
which helps us to plot a complete Hertzsprung-Russell (HR) diagram distribution, that can be easily compared with different isochrone models.

As a general rule, between two distinct stellar populations, the younger has larger metallicity. The extra content of heavy elements comes from the remains of older stars that already turned into supernovae, producing and releasing heavy elements to space that became later on part of the younger stars.
Similarly, bluer stars, generally more massive, 
are also younger than their redder counterparts, since their larger masses make their life shorter.
In conclusion, we can say that two stellar populations divided in colour or metallicity are also divided in age.
This fact could be one of the reasons why those studies that analysed different populations (e.g. references \cite{Sivertsson:2017rkp,Schutz:2017tfp,Buch:2018qdr} in Fig.~\ref{fig:rho_vs_years_recent}) obtained a different $\localDM$ for each of the populations.

\subsection{Differences in the methods}\label{subsec:diff-methods}
Even under common assumptions for the mass distribution of the Galaxy, the estimated value of $\localDM$ can be very sensitive to the method used in the analysis. 
For example, if a spread distribution of matter is present in the Milky Way forming a very thin disc
which is not accounted for in the study, 
the presence of this unaccounted matter would hardly affect the $\localDM$ estimated from an analysis of the rotation curve of the Galaxy, as long as the matter has a baryonic origin (the effect of a possible dark matter disc is discussed later in section~\ref{sec:dark-disc}). Nevertheless, this extra but not modelled thin disc will impact strongly the studies based on the vertical kinematics of local stars.
On a similar basis, small deviations in the density of baryons with respect to the assumed distribution could have a stronger effect in the estimated $\localDM$ when the studied volume is small, as it is generally the case in those analyses based on the local $z$-Jeans equation method. 
This means that 
a local study can give a resulting $\localDM$ closer to its true value than applying a global method, provided that the local distribution of baryons is well known or really well modelled. 
If the assumed local distribution of baryons is however far from its real shape, a global method performs better,
since a badly controlled systematic error might be added to the estimate coming from a local method.

Given the different approaches followed in the $\localDM$ estimates presented in Fig.~\ref{fig:rho_vs_years_recent}, the discrepancy in the central values and in the uncertainties of those results could be partly explained because of the differences in the analyses.
For instance, the three estimates using the vertical Jeans equation method previous to Gaia's observation (i.e. Refs.~\cite{McKee:2015hwa,Xia:2015agz,Sivertsson:2017rkp}) agree quite well with each other. These estimates are focused on a local volume that \emph{excludes} the Galactic plane (for example, \cite{Sivertsson:2017rkp} analysed stars within a vertical distance of $|z| = \text{515--1247}\,\mathrm{pc}$ for their $\alpha$-young population and $|z| = \text{634--2266}\,\mathrm{pc}$ for their $\alpha$-old population).
On the contrary, the three $\localDM$ studies that used Gaia's information (references~\cite{Schutz:2017tfp,Buch:2018qdr,Widmark:2018ylf}) present larger uncertainties and some of their estimates do not agree that well with the three older studies. One important difference is that the new estimates are concentrated on the local $|z| < 200\,\mathrm{pc}$, \emph{including} the Galactic plane.
Possibly, the differences in the results obtained in the three studies based on the vertical Jeans equation method that used Gaia's observations, and the three studies that did not, are related to the inclusion/exclusion of the Galactic plane, respectively, as well as the smaller volume covered in the works that included Gaia DR2 in their analyses.

There are several reasons why a smaller volume could lead to larger uncertainties. Generally, the smaller the volume under study, the more likely the analysis is affected by inhomogeneities in the local mass distribution and by the presence of local disequilibria.

One remark should be made about the $\localDM$ estimate from \cite{Widmark:2018ylf} shown in Fig.~\ref{fig:rho_vs_years_recent}. 
The authors of that work did not include an estimate of $\localDM$ in their study. 
Instead, they fitted the vertical total energy density distribution, regardless of its baryonic or dark matter origin, by studying Gaia's observations within a spherical shell of $100\,\mathrm{pc} < |z| < 200\,\mathrm{pc}$ from the position of the Earth in the Galaxy.
However, in their Fig.~4 they show a comparison of their determined total vertical energy density with respect to the vertical distribution of a typical baryonic model (the same that \cite{Schutz:2017tfp} used). 
The $\localDM$ estimate from \cite{Widmark:2018ylf} shown in our Fig.~\ref{fig:rho_vs_years_recent} corresponds to a direct comparison of the total energy density shown in their Fig.~4, at $z = 25\,\mathrm{pc}$,
and that of the same baryonic model used in \cite{Schutz:2017tfp}, interpreting the excess as produced by DM. 
However, we agree with the authors of that work that the excess could accommodate not only DM, but also a surplus of cold gas, which might be underestimated in typical models.

\subsection{Local disequilibria}\label{subsec:disequilibria}
Observations of asymmetries in the velocities and densities of stars in the Galactic disc show that the disc is experiencing vertical oscillations \cite{Antoja_2018}. A possible explanation for this feature is either the buckling of the Galactic bar \cite{Khoperskov:1811.09205} or the passage of a massive satellite \cite{Laporte:1808.00451}.

As we have already mentioned, disequilibria can affect $\localDM$ estimates, in particular those obtained from
local methods \cite{Banik:2016yqm}. This fact adds another possible explanation for the larger discrepancies shown in Fig.~\ref{fig:rho_vs_years_recent} among $\localDM$ estimates from analyses using a local method (based on the vertical Jeans equation) and global method determinations.

\subsection{New physics}\label{sec:dark-disc}
A clear way in which new physics can affect $\localDM$ estimates is 
through the hypothetical presence of 
a dark matter disc in the Milky Way. This possibility has been contemplated in several studies (see e.g. \cite{Read:2008fh,Purcell:2009yp,Fan:2013yva}). 
Although the presence of a dark disc is not ruled out, its mass and scale height are constrained to be small from Gaia's observations. This is the main reason why the studies of \cite{Schutz:2017tfp,Buch:2018qdr} analysed such a close environment to our location in the Galaxy.

One way in which a dark disc can affect $\localDM$ estimates works similarly to that of an unaccounted baryonic thin disc, as described at the beginning of Sec.~\ref{subsec:diff-methods}.
However, contrary to the case when the extra disc is made of baryons, a local method is more appropriate to study $\localDM$ in the presence of a disc composed of DM, provided that the real distribution of baryons is properly modelled.
A dark disc would show up in a study based on the vertical Jeans equation method as a larger value of the estimated $\localDM$.
Therefore, a local method could accommodate both contributions to $\localDM$, the one coming from the disc and the one from the halo.
Conversely, 
if the disc is sufficiently spread out along the Galactic plane 
(and given that the dark disc can not have a very large mass because most of the Galactic DM is present in the halo), 
its contribution to the rotation curve will be small and its presence might not affect at all the $\localDM$ estimate from the rotation curve method, that will acquire only the value of the contribution from the halo.

\subsection{Uncertainties in the distribution of baryons}
As we have pointed out before, the accuracy of a $\localDM$ estimate is subject to the accuracy of the relevant distribution of baryons. If the analysis is focused on the vertical movement of stars, it is the vertical baryonic distribution the one that matters most, while how important the radial distribution is will depend on the extension of the data.
On the contrary, if what is analysed is the rotation curve of the Galaxy, then the radial distribution becomes more relevant.
However, in analyses based on the rotation curve method what really matters is the mass enclosed at the probed distances from the Galactic centre.

\begin{figure}
\centering
\includegraphics[width=0.6\textwidth]{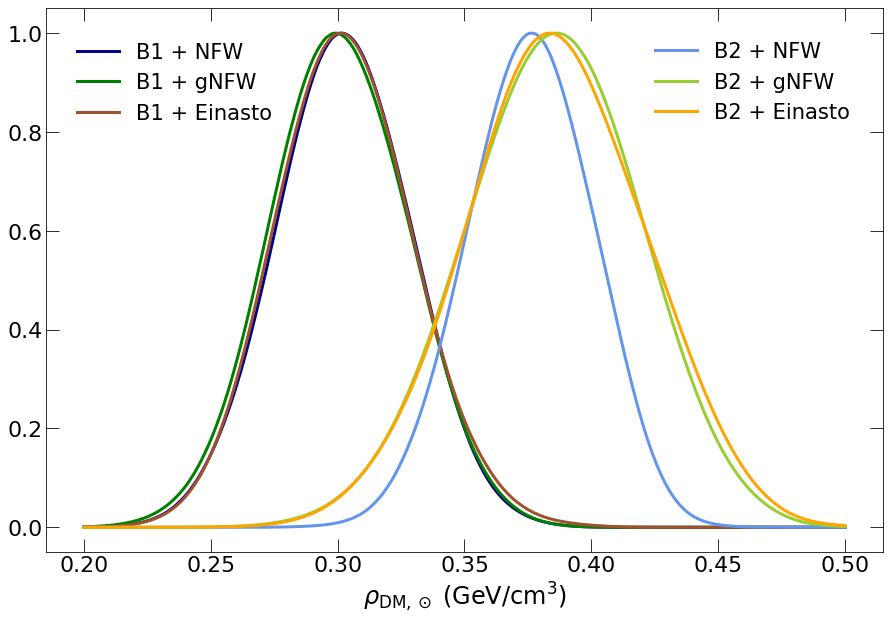}
\caption{Marginalised posterior distributions of the $\localDM$ estimates for the different cases analysed in \cite{deSalas:2019pee}. Two baryonic models are used: 
B1 and B2 (see the details in \cite{deSalas:2019pee}).
Three spherical dark halos are also considered (NFW, gNFW and Einasto). The main difference comes from the choice in the distribution of baryons, and the analyses of the baryonic B2 model prefer higher $\localDM$ values. Figure taken from \cite{deSalas:2019pee}. See the original reference for more details.
}
\label{fig:1D-posterior}
\end{figure}

To illustrate how much a change in the distribution of baryons can affect the estimated $\localDM$ value, in spite of the precision of the data, we show in Fig.~\ref{fig:1D-posterior} the marginalised posterior distributions of the $\localDM$ estimates from the different cases analysed in \cite{deSalas:2019pee}.
The accuracy of the circular velocity data used (obtained in \cite{Eilers:1810.09466} from observations of several surveys, including Gaia DR2) is of a few percent along most of the radial distance covered, from $R = 5\,\mathrm{kpc}$ to $R = 25\,\mathrm{kpc}$ from the Galactic centre.
However, despite such precise estimate of the circular velocity curve of the Milky Way, when this data is used to make an estimate of $\localDM$ the choice of the baryonic profile increases the associated uncertainties of the analysis.

\section{New approaches to estimate $\boldsymbol{\localDM}$}
Among the new ideas proposed in the literature that could help us obtain a better estimate of $\localDM$, there is one that has already given extraordinary results in its original field of application. As suggested by \cite{Ravi:2018vqd,Silverwood:2018qra}, the radial velocity method, that has proven so successful in the detection of exoplanets, could also be helpful in order to estimate $\localDM$.
The idea is to use the power of Doppler spectroscopy to follow the change in radial velocity directly caused by Galactic acceleration.
Since we are also part of the Galaxy, the followed stars need to be at a different radial location than our own in the Milky Way in order to distinguish their Galactic acceleration from the one felt at $R_\odot$.
This method has the advantage of directly sample the gravitational potential of the Galaxy, which reduces the amount of theoretical assumptions, so it is worth pursuing.
However, current precision of used spectrographs is not good enough, and 
future spectrographs will need to deal with many other effects that should be disentangled from the change in radial velocity caused by the acceleration of the Milky Way. For example, 
the presence of accompanying stars or orbiting planets would act as a background source.

\section{Conclusions}
Thanks to the observations from the Gaia satellite, at present we have very precise information about our Galaxy on hand.
However, recent estimates of $\localDM$ still rely on too simple Galactic models.
Working out a better model for the Galaxy is, nonetheless, a difficult task. 
There is an increasing number of publications moving into that direction, but important uncertainties remain in the distribution of baryons even at our local neighbourhood.
In addition, a better model of the Milky Way needs to be accompanied by a more complex analysis, putting different techniques at work at the same time, including several tracer populations and cross matching different observations to increase the precision of some stellar properties when required.

So far, if we consider typical assumptions such as an axisymmetric, steady-state Galaxy, and common baryonic models motivated by observations, the local dark matter density is constrained to be within $\localDM \simeq \text{0.3--0.4}\,\mathrm{GeV/cm^3}$ (as it was obtained in \cite{deSalas:2019pee} from the rotation curve of the Galaxy, a method that is expected to be less affected by local disequilibria
than local methods).
Smaller values would be difficult to explain, especially when we also take into account the somewhat larger estimates obtained in recent analyses based on the vertical Jeans equation method.
Larger values, on the other hand, are easily accommodated, since we could live in a particularly overdense region or the Galaxy could host an unexpected thin disc with either a baryonic or dark origin. In both such cases the rotation curve method could be blind to the extra content of local dark matter.

\section*{Acknowledgments}
Work supported by the Vetenskapsr{\aa}det (Swedish Research Council) through contract No. 638-2013-8993 and the Oskar Klein Centre for Cosmoparticle Physics.

\section*{References}

\end{document}